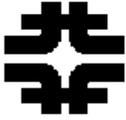

# Mass Resolution Corrections at the LHC

Dan Green
Fermilab
dgreen@fnal.gov

**Introduction**

The utility of jet spectroscopy at the LHC is compromised by the existence of multiple interactions within a bunch crossing. The energy deposits from these interactions at the design luminosity of the LHC may degrade the dijet mass resolution unless great care is taken. Energy clusters making up the jet can be required to have an energy flow with respect to the jet axis which resembles QCD [1]. In addition, subsidiary information such as the jet mass or the out of jet cone mass or transverse momentum can be deployed so as to alleviate the adverse effects of pileup [2].

A visual indication of the problem for jet spectroscopy is shown in Fig.1. Sequential Z events of 120 GeV mass decaying into light quark jets were generated and the CMS detector was simulated [3]. The resulting energy deposits at 1/5 design luminosity (LO) and design luminosity (HI) with tower sizes appropriate to the hadron calorimeter are shown. Clearly, at the HI, or design, luminosity adverse effects might be expected.

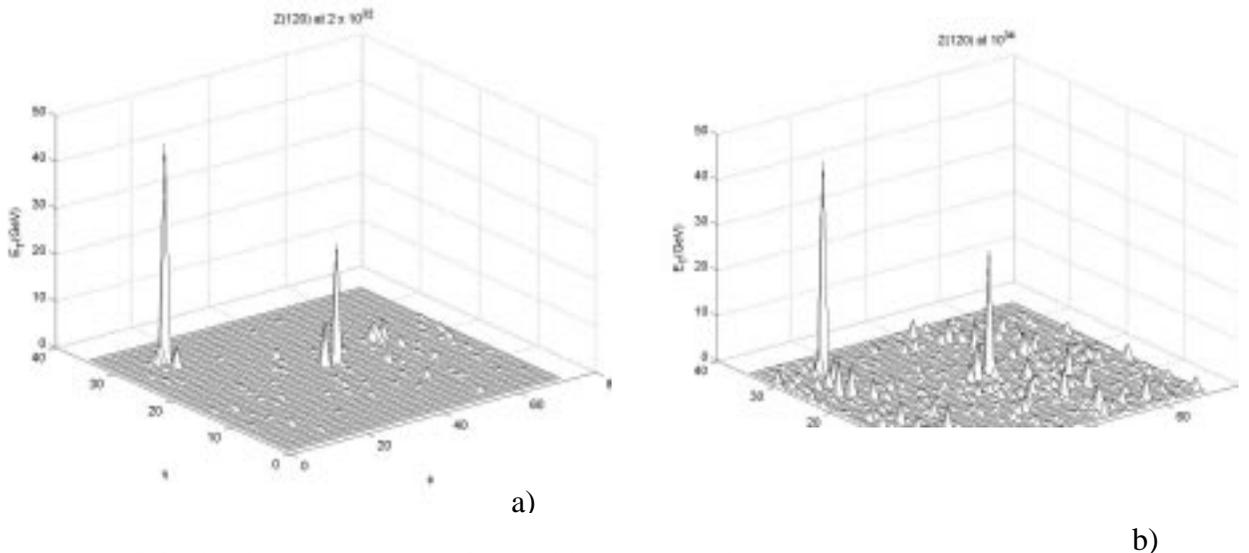

Figure 1: LEGO plot for dijet decays of a Z(120) event LHC bunch crossing at a luminosity of 2 x $10^{33}$ (LO), a), and $10^{34}$ /cm$^2$sec (HI), b). The segmentation is taken to be that of the CMS hadron calorimeter.

**Choice of Cone Size**

In the LO luminosity case there are about 330 distinct calorimeter clusters within the ten units of rapidity covered in this study. For the HI luminosity, the number is about 1100 clusters. Assuming 20 inelastic events per bunch crossing in the HI case and a flat rapidity plateau, each event has a density ~ 5 distinct clusters per unit of rapidity. There are ~ 8000 hadron calorimeter towers, so that the LO occupancy is ~ 2.4% and the HI occupancy is ~ 12 %.

Jets are defined to be a collimated deposit of transverse energy. This energy is collected within a cone of radius Rc centered on a seed tower, and the cone axis is then iterated. Clearly, a large cone radius allows the capture of ~ all the jet fragments. However, as seen in Fig.1, a large cone also admits many energy deposits unrelated to the jet due to random overlap with pileup events and the underlying event. The dijet mass, the jet mass, and the out of cone jet mass as a function of cone radius are shown in Fig.2 for Z(120) events at LO luminosity.

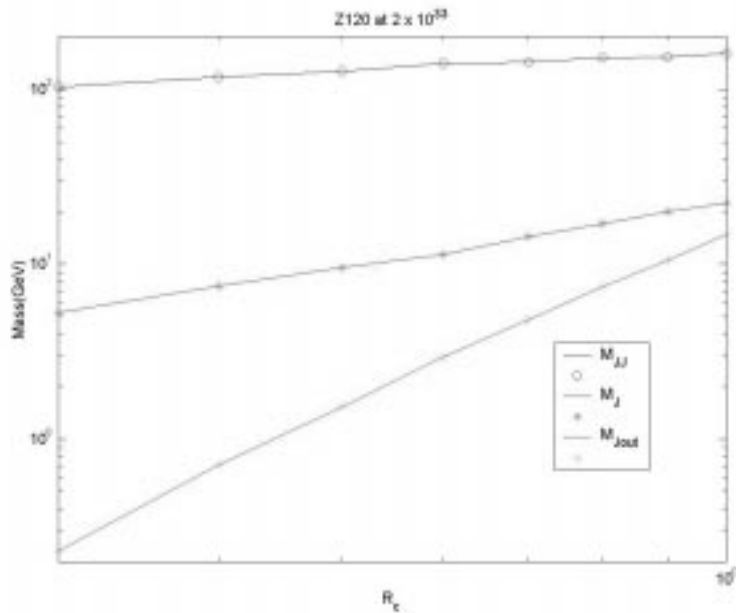

Figure 2: The dijet mass, the jet mass and the out of cone jet mass as a function of cone radius for Z(120) events at the LO luminosity operating point.

The out of cone mass is defined to be the mass of a cone centered on the jet in rapidity but at 90 degrees to the jet in azimuth. Clearly, the dijet mass changes a factor of two when the cone radius, Rc, varies from 0.3 to 1.0. The jet and out of cone jet masses vary even more rapidly. A medium value of Rc = 0.5 was chosen for the subsequent work.

Some representative mass and transverse momentum values are given in Table 1 for the four data sets studied here; Z(120) at LO and HI luminosity, and Z(700) at HI and LO luminosity

Table 1: Masses and Transverse Momenta Associated with Dijets (GeV units)

|  | Z(120) – LO | HI | Z(700) – LO | HI |
|---|---|---|---|---|
| $M_{Jout}$ | 1.45 | 6.1 | 1.8 | 6.2 |
| $M_J$ | 7.4 | 11.0 | 19.8 | 21.2 |
| $P_{TJout}$ | 5.0 | 18.1 | 6.6 | 18.8 |

The out of cone transverse momentum and out of cone mass are approximately independent of the Z' mass, because the underlying event and the pileup events are independent of the resonant mass. Both increase with luminosity, as expected. The jet masses are more weakly dependent on the luminosity, because they have major contributions from the real jet fragments.

The dijet mass distribution for Z(120) HI events is shown in Fig.3. The events with masses above 300 GeV are shown as overflows. The fractional mass resolution from a Gaussian fit to the data is about 25% at this luminosity, which indicates degradation due to the pile up energy contributions to the mass. The mean mass is 124 GeV as extracted from the fit to the distribution.

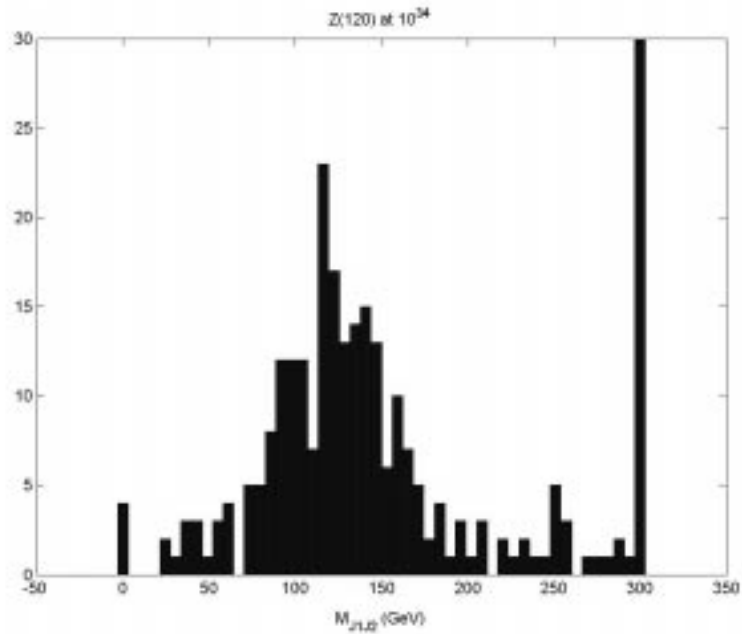

Figure 3: Dijet mass distribution for Z(120) events at HI luminosity using a cone radius of Rc = 0.5 and a seed tower threshold of 6 GeV. Events above 300 GeV are shown as overflows.

**Energy Flow within the Cone**

There is a characteristic difference between jet fragments and random pileup within the jet cone. The pile up contributions are approximately uniformly distributed per unit area in $(\eta,\phi)$ space. The inclusive transverse energy falls very rapidly at low momentum. Therefore, the background is expected to be uniformly distributed in dR, the distance from the jet axis. The distribution is also expected to be localized at small transverse momenta with respect to the total jet transverse momentum $z = \ln(x)$, $x = p_{Ti}/P_{TJ}$. In contrast, the jet is expected to have a "core" of high transverse momentum near the jet axis and softer fragments at lower transverse momentum which occur at larger dR values. Data for Z(120) HI events is shown in Fig.4.

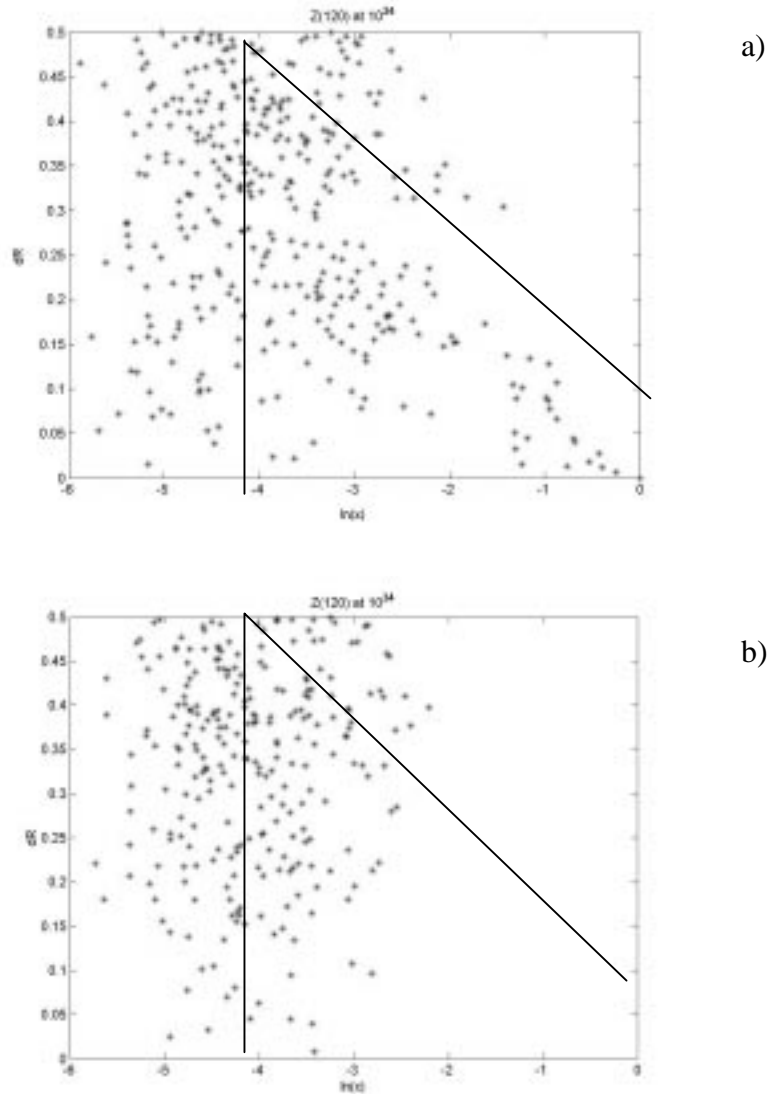

Figure 4: Energy flow within a jet, a), and within an out of jet cone, b). The points populate the radial distance from the jet axis, dR, and the fraction of the jet transverse momentum, $\ln(x_i = p_{Ti}/P_{TJ})$. The lines indicate the cuts made when allowing calorimeter clusters into the jet momentum sum.

It is clear that the cuts which are applied discriminate between pileup and real jet fragments. The jet "core" which has z > -2.5 is retained along with a fraction of the softer fragments. These fragments are selected with energy-angle ordering in mind. Large angle fragments are expected to be soft. All fragments with less than about 2% (z < -4) of the jet momentum are removed. This cut will potentially limit very precise jet spectroscopy.

The dijet mass distribution for Z(120) events at HI luminosity is shown in Fig.5 for jets where the cuts indicated in Fig.4 have been applied. Comparing to Fig.3, a clear improvement in the mass resolution has been made. A fit to the mass distribution yields a 17% fractional mass resolution compared to the previous 25%.

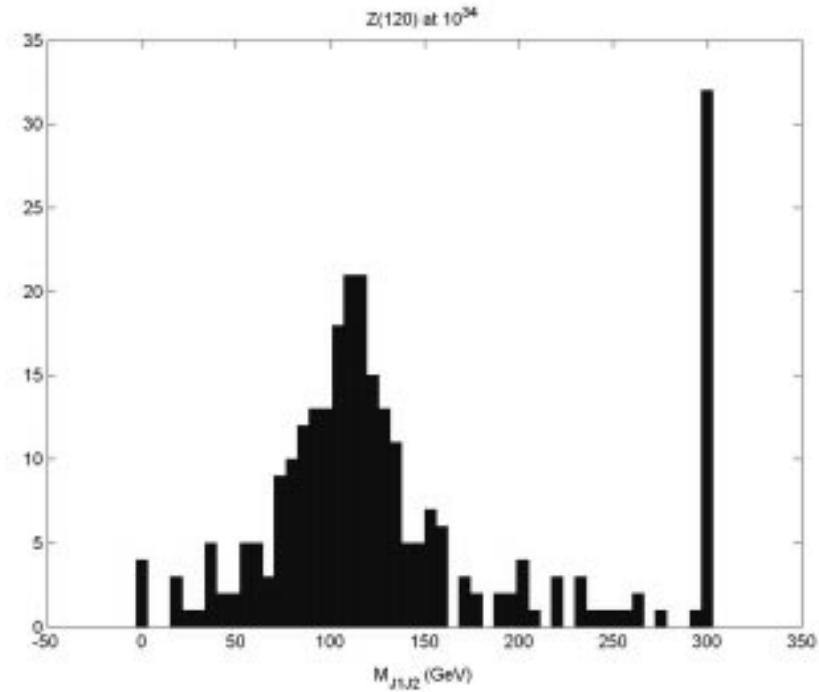

Figure 5: Dijet mass distribution for Z(120) events at HI luminosity using a cone radius of Rc = 0.5 and a seed tower threshold of 6 GeV. Events above 300 GeV are shown as overflows. The cuts on clusters indicated in Fig.4 have been applied.

**Dijet Mass Corrections Using Jet Masses**

The fragmentation process leads to jets possessing a mass if the fragments are taken to be massless. Nevertheless, the fractional dijet mass resolution depends very weakly on whether or not the jets are taken to be massless. Indeed, that resolution depends only on the 3 dimensional momentum vectors of the two jets. Assuming that the 2 jets are back-to-back in azimuth, the dijet mass depends on only 4 variables, as seen in Eq.1.

$$M_{12} \sim 4P_{T1}P_{T2}\cosh^2((y_1-y_2)/2) \qquad (1)$$

The mass of the jet depends on the distribution of the jet fragments with respect to the axis and the transverse momentum of the fragment, as shown in Eq.2. Nevertheless, the jet mass is proportional to the jet transverse momentum, and therefore (see Eq.1), one might attempt to use the jet mass to correct the dijet mass for pile up, radiation, and energy mis-measurement.

$$M_1^2 = P_{T1}\sum_i p_{ti}dR_i^2$$
$$dR_i = \sqrt{d\eta_i^2 + d\phi_i^2} \qquad (2)$$

The variables which are available beyond those needed to compute the dijet mass are the two jet masses, and the mass and transverse momentum in the out of jet cones for the two jets. Those variables were examined and only the jet masses were found to be correlated with the dijet mass. Therefore, a correction expressed in Eq.3 was defined with which to correct the dijet mass. Clearly, the functional form of the correction is driven by Eq.1. The constant terms are defined so as not to pull the mean mass around, at least for the Z(120) HI luminosity sample. The constant a is a free variable which is chosen to optimize the dijet fractional mass resolution.

$$M_{12} - a[\sqrt{M_1 M_2}\cosh((y_1-y_2)/2) - \sqrt{<P_{T1out}><P_{T2out}>}] \qquad (3)$$

The correlation between the functional form of the jet masses and the dijet mass for Z(120) LO , Z(120) HI, Z(700) LO and Z(700) HI events is shown in Fig.6. The jet rapidities are used in the correction term because they correct for the fact that the p-p C.M. is not the jet-jet C.M. frame. The jet masses themselves reflect jet fragmentation, but also pile up contributions, final state radiation, and calorimeter cluster energy mis-measurements.

The dijet mass distribution for events corrected by means of Eq.3 for Z(120) HI luminosity is shown in Fig.7. Comparing to Fig.5, the very large masses (overflows in the histogram) have been reduced, as have the low dijet masses. In addition, the main distribution has been sharpened, as is expected given the correlation between the jet masses and the dijet mass displayed in Fig.6.

The results of Gaussian fits for uncorrected dijet masses, (z,dR) corrected masses, and (z,dR) plus jet mass corrected dijet masses for the Z(120) HI sample are given in Table.2. The fractional mass distribution decreases from 25% to 17% to 13% as the cuts and corrections are made. Clearly, good progress can be made in improving the jet masses using techniques developed for operation at the high luminosities which will be encountered at the LHC.

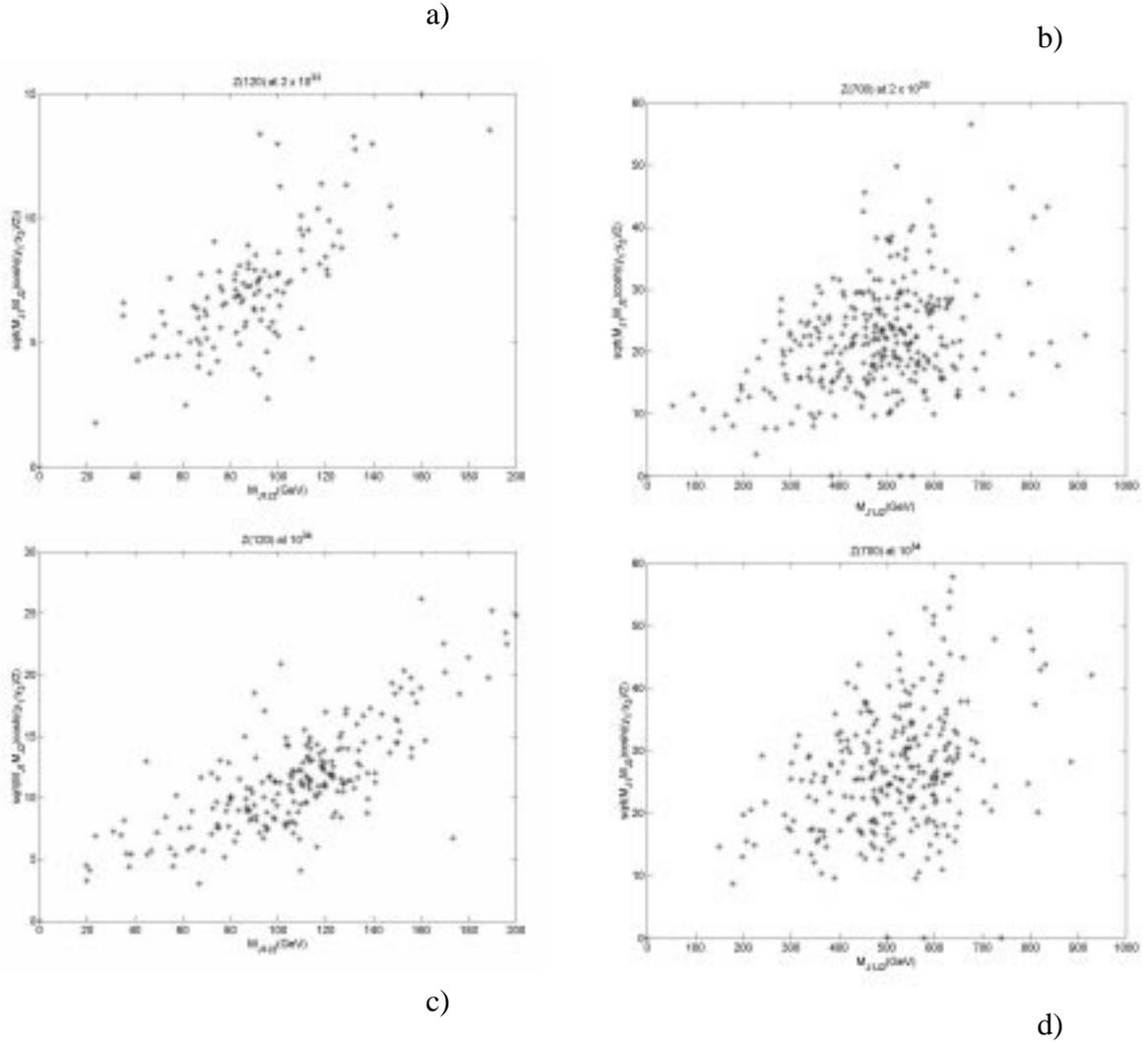

Figure 6: Correlation between the dijet mass calculated assuming massless jets and the jet masses, specifically $\sqrt{M_1 M_2} \cosh((y_1 - y_2)/2)$. The scatter-plots are for Z(120) LO, a), Z(120) HI, b), Z(700) LO, c), and Z(700) HI, d). The different mass resonances and the different luminosity levels all show a correlation.

Table 2: Gaussian Fits to the Dijet Mass Distribution (GeV units)

|              | Mean     | σ         | $\chi^2$/DOF |
|--------------|----------|-----------|--------------|
| Z(120) HI    | 124+-4.0 | 31.3 +-6.0 | 12.6/7      |
| (ln(x),dR) Cuts | 111+-2.3 | 19.4 +-2.2 | 5.4/7   |
| Corrected    | 138+-1.7 | 17.9 +-2.0 | 2.6/7       |

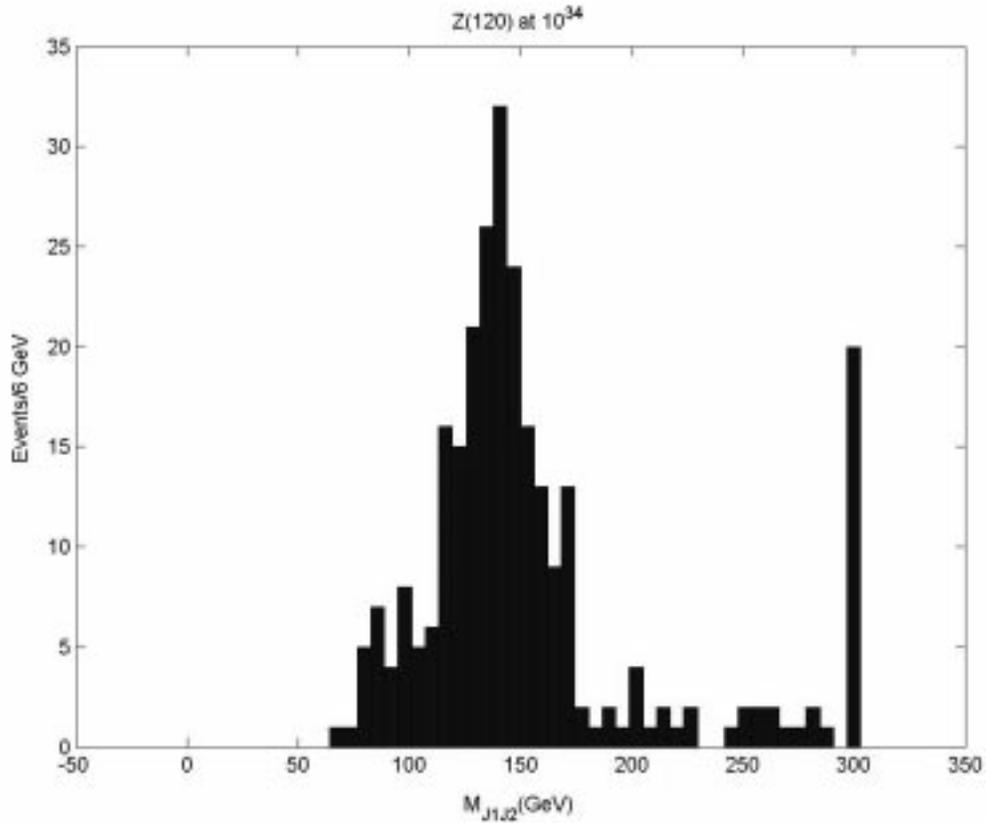

Figure 7: Dijet mass distribution for Z(120) events at HI luminosity using a cone radius of Rc = 0.5 and a seed tower threshold of 6 GeV. Events above 300 GeV are shown as overflows. The cuts on clusters indicated in Fig.4 have been applied. In addition, the mass correction indicated in Eq.3 has been applied.

**Summary**

The mass resolution for Z(120) events was studied using cuts in the (z,dR) plane to sort on energy clusters within a jet cone and by using a correction which utilized the independent subsidiary variables, the jet masses. Both methods improved the mass resolution. By applying the two sequentially, a factor of almost two improvement was achieved.

The fact that Z(120) LO luminosity and Z(700) HI and LO luminosity data also showed correlation between the dijet mass and the jet masses, implies that these techniques are of general applicability.

It can be expected that these improvements mentioned above will be independent of the "energy flow" techniques which can also me used to use the momentum measurements of charged tracks to replace the calorimeter energy measurements of matched clusters of deposited energy [4].